\documentclass{elsart}

\usepackage{epsfig, color, amsmath, amssymb, gastex, graphicx}

\begin{document}

\newcommand{\fsgd}{F_\sigma \cap G_\delta}
\newcommand{\nats}{\mathbb{N}}
\newcommand{\reals}{\mathbb{R}}
\newcommand{\ints}{\mathbb{Z}}
\newcommand{\infi}{\mbox{\it inf\/}}
\newcommand{\point}{\star}
\newcommand{\lep}{\ell ep}
\newcommand{\cop}{cop}
\newcommand{\fcop}{\mbox{\bf cop}}
\newcommand{\flep}{\mbox{\bf lep}}
\newcommand{\gsp}{gsp}
\newcommand{\ogsp}{{\neg{\gsp}}}
\newcommand{\losp}{\ell osp}
\newcommand{\olosp}{{\neg{\losp}}}
\newcommand{\boolset}{\{0,1\}}
\newcommand{\confacc}{\mbox{\it accept}}

\newcommand{\TS}{\mathcal{T}}
\newcommand{\ETS}{\mathit{ETS}}
\newcommand{\PTS}{\mathcal{P}}
\newcommand{\BTS}{\mathcal{B}}
\newcommand{\ACC}{\mathit{ACC}}
\newcommand{\REL}{\mathit{REL}}
\newcommand{\trace}{\mathit{Tr}}
\newcommand{\inff}{\mathit{inf}}
\newcommand{\id}{\mathit{id}}
\newcommand{\liv}{\mathcal{L}}
\newcommand{\sav}{\mathcal{R}}

\begin{frontmatter}

\title{A Framework To Handle Linear Temporal Properties in ($\omega$-)Regular Model Checking \thanksref{title:also}}
\thanks[title:also]{The present article is an extended version of a paper
which appears in the Proceedings of \cite{BLW04b}.}

\author[Paris]{Ahmed Bouajjani},
        \ead{abou@liafa.jussieu.fr}
\author[Rennes]{Axel Legay},
        \ead{alegay@irisa.fr} 
\author[Liege]{Pierre Wolper},
        \ead{pw@montefiore.ulg.ac.be}

\address[Paris]{
        LIAFA~-~Universit\'{e}~Paris~7\\
        175, rue du chevaleret\\
        Paris, France}
\address[Rennes]{
        Universit\'e de Rennes 1\\
        Institut d'informatique INRIA\\
        Rennes, France}
\address[Liege]{
        Universit\'e de Li\`ege\\
        Institut Montefiore, B28\\
        Li\`ege, Belgium}

\begin{abstract}
Since the topic emerged several years ago, work on regular model
checking has mostly been devoted to the verification of state
reachability and safety properties. Though it was known that linear
temporal properties could also be checked within this framework,
little has been done about working out the corresponding details. This
paper addresses this issue in the context of regular model checking
based on the encoding of states by finite or infinite words. It works
out the exact constructions to be used in both cases, and proposes a
partial solution to the problem resulting from the fact that infinite
computations of unbounded configurations might never contain the same
configuration twice, thus making cycle detection problematic.
\end{abstract}

\end{frontmatter}

\newcommand{\proved}{\hfill $\Box$}
\newcommand{\tuple}[1]{\left({#1}\right)}
\newcommand{\set}[1]{\left\{#1\right\}}
\newcommand{\setcomp}[2]{\left\{#1|\;#2\right\}}

\newcommand{\alphabet}{\Sigma}
\newcommand{\eqalphabet}[1]{\Sigma^{\bullet}(#1)}
\newcommand{\arity}{\rho} 
\newcommand{\vars}{X}
\newcommand{\tree}{\mathit{Tr}}
\newcommand{\treestr}{S}
\newcommand{\labeling}{\lambda}
\newcommand{\ctreestr}{\treestr_{\context}}
\newcommand{\clabeling}{\labeling_{\context}}
\newcommand{\treetuple}{\tuple{\treestr,\labeling}}
\newcommand{\fsymb}{f}
\newcommand{\tnode}{s}
\newcommand{\cnode}{\tnode_c}
\newcommand{\automaton}{A}
\newcommand{\transducer}{T}
\newcommand{\states}{Q}
\newcommand{\finalstates}{F}
\newcommand{\transitionrel}{\triangle}
\newcommand{\transitionrule}[3]{\left(#1\right)\stackrel{#2}{\longrightarrow}{#3}}
\newcommand{\emptytransitionrule}[2]{\stackrel{#1}{\longrightarrow}{#2}}
\newcommand{\automatontuple}{\tuple{\states,\alphabet,\transitionrel,\finalstates}}
\newcommand{\transducertuple}{\tuple{Q,\Sigma^\bullet(2), \triangle,F}}
\newcommand{\stransducer}{S}
\newcommand{\sattransducer}{S_{{\it SAT}}}
\newcommand{\sstates}{Q_S}
\newcommand{\sfinalstates}{F_S}
\newcommand{\stransitionrel}{\triangle_S}
\newcommand{\stransducertuple}{\tuple{\sstates,\Sigma^\bullet(2),\stransitionrel}, \sfinalstates}
\newcommand{\htransducer}{H}
\newcommand{\hstates}{Q_H}
\newcommand{\hfinalstates}{F_H}
\newcommand{\htransitionrel}{\triangle_H}
\newcommand{\htransducertuple}{\tuple{\hstates, \Sigma^\bullet(2),\htransitionrel, \hfinalstates}}
\newcommand{\str}{w}
\newcommand{\state}{q}
\newcommand{\cpystate}{\state_{\it cpy}}
\newcommand{\idmstate}{\state_{\it idm}}
\newcommand{\hstate}{\str}
\newcommand{\run}{r}
\newcommand{\runsto}[2]{\stackrel{#1}{\Longrightarrow}_{#2}}
\newcommand{\languageof}{L}
\newcommand{\lang}{K}
\newcommand{\emptystring}{\epsilon}
\newcommand{\relation}{R}
\newcommand{\relationof}[1]{\left[#1\right]}
\newcommand{\rel}{R}
\newcommand{\product}{\times}
\newcommand{\abs}[2]{{#1}|_{#2}}
\newcommand{\disj}{\cup}
\newcommand{\composition}{\circ}
\newcommand{\sizeof}[1]{|#1|}
\newcommand{\regulartransitionrel}[2]{{#1}\stackrel{\left(#2\right)}{\longrightarrow}}

\newcommand{\word}{w}
\newcommand{\words}{W}
\newcommand{\length}[1]{|#1|}
\newcommand{\re}{\phi}
\newcommand{\hole}{\Box}
\newcommand{\suff}{\operatorname{suff}}
\newcommand{\pref}{\operatorname{pref}}
\newcommand{\saturate}[2]{\lceil#1\rceil_{#2}}

\newcommand{\satured}[1]{\lceil#1\rceil}
\newcommand{\context}{C}
\newcommand{\upsim}{\preccurlyeq_{up}}
\newcommand{\downsim}{\preccurlyeq_{down}}

\newcommand{\qpref}{\state_{\it pref}}
\newcommand{\qsuff}{\state_{\it suff}}

\newcommand{\qprefs}{\states_{\it pref}}
\newcommand{\qsuffs}{\states_{\it suff}}

\newcommand{\structproduct}{\otimes}
\newcommand{\dtuple}{\tuple{Q_T, F_T, \delta_T}}

\section{Introduction}

At the heart of all the techniques that have been proposed for
exploring infinite state spaces, is a symbolic representation that
can finitely represent infinite sets of states. In early work on the
subject, this representation was domain specific, for example linear
constraints for sets of real vectors. For several years now, the idea
that a generic finite-automaton based representation could be used in
many settings has gained ground, starting with systems manipulating
queues and integers~\cite{WB95,BEM97,BRW98}, then moving to
parametric systems~\cite{KMMPS97}, and, recently, reaching systems
using real variables~\cite{BJW01,BHJ03}. 

Beyond the necessary symbolic representation, there is also a need to
``accelerate'' the search through the state space in order to reach,
in a finite amount of time, states at unbounded depths. In
acceleration techniques, the move has again been from the specific to
the generic, the latter approach being often referred to as regular
model checking. In ($\omega$-)regular model checking (see
e.g.~\cite{BJNT00,DLS02,BLW04a}), the transition relation is
represented by a finite-state transducer and acceleration techniques
aim at computing the iterative closure of this transducer
algorithmically, though necessarily foregoing totality or preciseness,
or even both. The advantages of using a generic technique are of
course that there is only one method to implement independently of the
domain considered, that multidomain situations can potentially be
handled transparently, and that the scope of the technique can include
cases not handled by specific approaches. Beyond these concrete
arguments, one should not forget the elegance of the generic approach,
which can be viewed as an indication of its potential, thus justifying
a thorough investigation.  

However, computing reachable states is not quite model-checking. For
reachability properties model checking can be reduced to a state
reachability problem, but for properties that include a linear
temporal component, the best that can be done is to reduce the
model-checking problem to emptiness of a B\"uchi
automaton~\cite{VW86}, which represents all the executions of the
system that do not satisfy the property. If this automaton is empty,
then the system satisfies the property, else the property is not
satisfied. In this framework, one thus has to check for repeated
reachability rather than reachability.

In this paper, we consider the specification and the verification of
linear temporal properties in the ($\omega$-)regular model checking
framework{\footnote{In the rest of the paper, we use
    ``($\omega$-)regular model checking'' to denote either ``regular
    model checking'' or ``$\omega-$regular model checking'', depending
    on whether states are encoded by finite or infinite words.}}. The
objective of the paper is to provide generic analysis techniques
covering various classes of systems that can be encoded in this
framework.

We fully worked out how to augment the transducer representing the
system transitions in order to obtain a transducer encoding the
B\"uchi automaton resulting from combining the system with the
property. Once the transition relation of the B\"uchi automaton has
been obtained, checking the automaton for nonemptiness is done by
computing the iterative closure of this relation, finding nontrivial
cycles between states, and finally checking for the reachability of
states appearing in such cycles. When dealing with systems where the
number of successors of each state is bounded, an accepting execution
of the B\"uchi automaton will always contain the same state twice and
hence an identifiable cycle. However, when dealing with states whose
length can grow or that are infinite, there might very well be an
accepting computation of the B\"uchi automaton in which the same state
never appears twice.

To cope with this, we look for states that are not necessarily
identical, but such that one entails the other in the sense that
any execution possible from one is also possible from the other. The
exact notion of entailment we use is simulation.  For that, we compute
symbolically the greatest simulation relation on the states of
the system.

The nice twist is that the computation of the symbolic representation
of the simulation relation is in fact, the computation of the limit of
a sequence of finite-state automata, for which the acceleration
techniques introduced in \cite{BLW03,BLW04a,Leg07} can be
used. However, there are also several cases where this computation
converges after a finite number of steps, which has the added
advantage of guaranteeing that the induced simulation equivalence
relation partitions the set of configurations in a finite number of
classes, and hence that existing accepting computations will
necessarily be found, which might not be the case when the number of
simulation equivalence classes is infinite.\\
\newline

{\bf Structure of the paper.}  The paper is structured as follows. In
Section \ref{automata}, we recall the elementary definitions on
automata theory that will be used throughout the rest of the
paper. Section \ref{paper-model} presents the ($\omega$-)regular model
checking framework as well as a methodology to reason about infinite
executions. In Sections \ref{temporal-rmc1}, \ref{last-thesis},
\ref{temporal-rmc2}, and \ref{bool-liveness}, the verification of
several classes of linear temporal properties in the
($\omega$-)regular model checking framework is considered. Finally,
Sections \ref{related} and \ref{conclu} conclude the paper with a
comparison with other works on the same topic and several directions
for future research, respectively.

\section{Background on Automataa Theory}
\label{automata}
In this section, we introduce several notations, concepts, and
definitions that will be used throughout the rest of this paper. The
set of natural numbers is denoted by $\nats$, and $\nats_0$ is used
for $\nats\setminus {\lbrace}0{\rbrace}$.

\subsection{Relations}

Consider a set $S$, a set $S_1\,\subseteq\,S$, and two
binary{\footnote{The term ``binary'' will be dropped in the rest of
    the paper.}} relations $R_1, R_2\,\subseteq\,S\times S$. The
identity relation on $S$, denoted $R_\id^S$ (or $R_{id}$ when $S$ is
clear from the context) is the set ${\lbrace}(s,s)|s\in
S{\rbrace}$. The {\em image} of $S_1$ by $R_1$, denoted $R_1(S_1)$, is
the set ${\lbrace}s'\in S_1\mid (\exists s\in S_1)( (s,s')\in
R_1){\rbrace}$. The {\em composition} of $R_1$ with $R_2$, denoted
$R_2\circ R_1$, is the set ${\lbrace}(s,s')\mid (\exists
s'')((s,s'')\in R_1\wedge (s'',s')\in R_2){\rbrace}$. The $i$th {\em
  power} of $R_1$ ($i\in \nats_0$), denoted $R_1^i$, is the relation
obtained by composing $R_1$ with itself $i$ times. The {\em
  zero-power} of $R_1$, denoted $R_1^0$, corresponds to the identity
relation. The {\em transitive closure} of $R_1$, denoted $R_1^+$, is
given by $\bigcup_{i=1}^{i=+\infty}R_1^i$, its {\em reflexive
  transitive closure}, denoted $R^*$, is given by $R_1^+\cup
R_{\id}^S$. The {\em domain} of $R_1$, denoted $\it{Dom}(R_1)$, is
given by ${\lbrace}s\in S\mid (\exists s'\in S)((s,s')\in
R_1){\rbrace}$.

\subsection{Words and Languages}
\label{word-language}

An {\em alphabet} is a (nonempty) finite set of distinct symbols.  A
{\em finite word\/} of length $n$ over an alphabet $\Sigma$ is a
mapping $w:{\lbrace}0,{\dots},n-1{\rbrace}{\rightarrow}\Sigma$. An
{\em infinite word \/}, also called $\omega-$word, over $\Sigma$ is a
mapping $w:{\nats}{\rightarrow}\Sigma$. We denote by the term {\em
  word} either a finite word or an infinite word, depending on the
context. The {\em length} of the finite word $w$ is denoted by
$|w|$. A finite word $w$ of length $n$ is often represented by
$w=w(0){\cdots}w(n-1)$. An infinite word $w$ is often represented by
$w(0)w(1){\cdots}$ . The sets of finite and infinite words over
$\Sigma$ are denoted by $\Sigma^*$ and by $\Sigma^{\omega}$,
respectively. We define $\Sigma^{\infty}=\Sigma^*\cup
\Sigma^{\omega}$. A {\em finite-word (respectively infinite-word)
  language} over $\Sigma$ is a (possibly infinite) set of finite
(respectively, infinite) words over $\Sigma$. Consider $L_1$ and
$L_2$, two finite-word (resp. infinite-word) languages. The {\em
  union} of $L_1$ and $L_2$, denoted $L_1\cup L_2$, is the language
that contains all the words that belong either to $L_1$ or to
$L_2$. The {\em intersection} of $L_1$ and $L_2$, denoted $L_1\cap
L_2$, is the language that contains all the words that belong to both
$L_1$ and $L_2$. The {\em complement} of $L_1$, denoted
$\overline{L_1}$ is the language that contains all the words over
$\Sigma$ that do not belong to $L_1$.\\
\newline
We alos introduce {\em synchronous product} and {\em projection},
which are two operations needed to define relations between languages.

\begin{defn}
\label{definition-cartesian-product}
Consider $L_1$ and $L_2$ two languages over $\Sigma$.
\begin{itemize}
\item
If $L_1$ and $L_2$ are finite-word languages, the synchronous product
$L_1\bar{{\times}} L_2$ of $L_1$ and $L_2$ is defined as follows
\begin{center}
$L_1\bar{{\times}}
L_2={\lbrace}(w(0),w(0)'){\dots}(w(n),w(n)')\mid$\\$
w=w(0)w(1){\dots}w(n)\in L_1\,{\wedge}\,w'=w(0)'w(1)'{\dots}w(n)'\in
L_2{\rbrace}$.
\end{center} 
\item
If $L_1$ and $L_2$ are $\omega$-languages, the synchronous product
$L_1\bar{{\times}} L_2$ of $L_1$ and $L_2$ is defined as follows
\begin{center}
$L_1\bar{{\times}}
L_2={\lbrace}(w(0),w(0)')(w(1),w(1)'){\cdots}\mid$\\$
w=w(0)w(1){\dots}\in L_1\,{\wedge}\,w'=w(0)'w(1)'{\cdots}\in
L_2{\rbrace}$.
\end{center} 
\end{itemize}
The language $L_1\bar{{\times}} L_2$ is defined over the alphabet
$\Sigma^2$.
\end{defn}

\noindent
Definition \ref{definition-cartesian-product} directly generalizes to
synchronous products of more than two languages. Given two finite
(respectively, infinite) words $w_1, w_2$ (with $|w_1|=|w_2|$ if the
words are finite) and two languages $L_1$ and $L_2$ with
$L_1={\lbrace}w_1{\rbrace}$ and $L_2={\lbrace}w_2{\rbrace}$, we use
$w_1{\bar{\times}}w_2$ to denote the {\em unique} word in
$L_1{\bar{\times}}L_2$.

\begin{defn}
\label{def-projection}
Suppose $L$ a language over the alphabet $\Sigma^n$ and a natural
$1\,{\leq}\,i\,{\leq}n$. The projection of $L$ on all its components
except component $i$, denoted $\Pi_{\not= i}(L)$, is the language $L'$
such that
\begin{center}
$\Pi_{\not=i }(L)={\lbrace}w_1\bar{\times}\dots\bar{\times}w_{i-1}\bar{\times}w_{i+1}\bar{\times}\dots
\bar{\times} w_n \mid$\\$ (\exists
w_i)(w_1\bar{\times}\dots\bar{\times}w_{i-1}\bar{\times}w_i\bar{\times}w_{i+1}\bar{\times}\dots
\bar{\times} w_n\in L){\rbrace}$.
\end{center}
\end{defn}

\subsection{Automata}

\begin{defn}
\label{finite-word-automaton}
An automaton over $\Sigma$ is a tuple $A=(Q,\Sigma,Q_0,\triangle,F)$,
where
\begin{itemize}
\item
$Q$ is a finite set of {\em states\/},
\item
$\Sigma$ is a {\em finite} alphabet,
\item
$Q_0\,\subseteq\,Q$ is the set of {\em initial states\/},
\item
$\triangle\,\subseteq\, Q\times \Sigma \times Q$ is a finite {\em transition
relation\/}, and
\item
$F\,\subseteq\,Q$ is the set of accepting states (the states in
$Q\setminus F$ are the {\em nonaccepting} states).
\end{itemize}
\end{defn}

Let $A=(Q,\Sigma,Q_0,\triangle,F)$ be an automaton and $a\in
\Sigma$. If $(q_1,a,q_2)\in \triangle$, then we say that there is a
{\em transition} from $q_1$ (the {\em origin}) to $q_2$ (the {\em
  destination}) labeled by $a$. We sometimes abuse the notations, and
write $q_2\in \triangle(q_1,a)$ instead of $(q_1,a,q_2)\in
\triangle$. Two transitions $(q_1,a,q_2), (q_3,b,q_4)\in \triangle$
are {\em consecutive} if $q_2=q_3$. Given two states $q,q'\in Q$ and a
finite word $w\in \Sigma^*$, we write $(q,w,q')\in \triangle^*$ if
there exist states $q_0,\dots,q_{n}$ and $w(0),\dots,w(n-1)\in \Sigma$
such that $q_0=q$, $q_{n}=q'$, $w=w(0)w(1)\cdots w(n-1)$, and
$(q_i,w(i),q_{i+1})\in \triangle$ for all $0\,{\leq}\,i<n-1$. Given
two states $q, q'\in Q$, we say that the state $q'$ is {\em reachable}
from $q$ in $A$ if $(q,a,q')\in \triangle^*$. The automaton $A$ is
{\em complete} if for each state $q\in Q$ and symbol $a\in \Sigma$,
there exists at least one state $q'\in Q$ such that $(q,a,q')\in
\triangle$. An automaton can easily be completed by adding an
extra nonaccepting state.\\
\newline A {\em finite run} of $A$ on a finite word
$w:{\lbrace}0,{\dots},n-1{\rbrace}{\rightarrow}\Sigma$ is a labeling
$\rho:{\lbrace}0,{\dots},n{\rbrace}{\rightarrow}Q$ such that
$\rho(0)\in Q_0$, and
$(\forall{0\,{\leq}\,i\,{\leq}\,n-1})((\rho(i),w(i),\rho(i+1))\in
\triangle)$. A finite run $\rho$ is {\em accepting} for $w$ if
$\rho(n)\in F$. An {\em infinite run} of $A$ on an infinite word
$w:{\nats}{\rightarrow}\Sigma$ is a labeling
$\rho:\nats{\rightarrow}Q$ such that $\rho(0)\in Q_0$, and
$(\forall{0\,{\leq}\,i})((\rho(i),w(i),\rho(i+1))\in \triangle)$. An
infinite run $\rho$ is {\em accepting} for $w$ if $\infi(\rho)\cap F
\neq \emptyset$, where $\infi(\rho)$ is the set of states that are
visited infinitely often by $\rho$.\\
\newline We distinguish between {\em finite-word automata} that are
automata accepting finite words, and {\em B\"uchi automata} that are
automata accepting infinite words. A finite-word automaton accepts a
finite word $w$ if there exists an accepting finite run on $w$ in
this automaton. A B\"uchi automaton accepts an infinite word $w$ if
there exists an accepting infinite run on $w$ in this automaton. The
set of words accepted by $A$ is the {\em language accepted by $A$},
and is denoted $L(A)$. Any language that can be represented by a
finite-word (respectively, B\"uchi) automaton is said to be {\em
  regular}
(respectively, {\em $\omega$-regular}).\\
\newline The automaton $A$ may behave nondeterministicaly on an input
word, since it may have many initial states and the transition
relation may specify many possible transitions for each state and
symbol. If $|Q_0|=1$ and for all state $q_1\in Q$ and symbol $a\in
\Sigma$ there is at most one state $q_2\in Q$ such that
$(q_1,a,q_2)\in \triangle$, then $A$ is {\em deterministic}. In order
to emphasize this property, a deterministic automaton is denoted as a
tuple $(Q, \Sigma, q_0,\delta,F)$, where $q_0$ is the unique initial
state and $\delta:Q\times \Sigma~{\rightarrow}~Q$ is a partial
function deduced from the transition relation by setting
$\delta(q_1,a)=q_2$ if $(q_1,a,q_2)\in \triangle$. Operations on
languages directly translate
to operations on automata, and so do the notations.\\
\newline One can decide weither the language accepted by a finite-word
or a B\"uchi automaton is empty or not. It is also known that
finite-word automata are closed under determinization,
complementation, union, projection, and
intersection\,\cite{Hop71}. Moreover, finite-word automata admit a
minimal form, which is unique up to
isomorphism\,\cite{Hop71}.\\
\newline Though the union, intersection, synchronous product, and
projection of B\"uchi automata can be computed efficiently, the
complementation operation requires intricate algorithms that not only
are worst-case exponential, but are also hard to implement and
optimize (see \cite{Var07a} for a survey). The core problem is that
there are B\"uchi automata that do not admit a deterministic/minimal
form. To working with infinite-word automata that do own the same
properties as finite-word automata, we will restrict ourselves to {\em
  weak\/} automata~\cite{MSS86} defined hereafter.

\begin{defn}
For a B\"uchi automaton $A=(\Sigma,Q,q_0,\delta,F)$ to be weak, there
has to be partition of its state set $Q$ into disjoint subsets $Q_1,
\ldots, Q_m$ such that for each of the $Q_i$, either $Q_i \subseteq
F$, or $Q_i \cap F = \emptyset$, and there is a partial order $\leq$
on the sets $Q_1, \ldots, Q_m$ such that for every $q \in Q_i$ and
$q'\in Q_j$ for which, for some $a \in \Sigma$, $q'\in \delta(q,a)$
($q'=\delta(q,a)$ in the deterministic case), $Q_j \leq Q_i$.
\end{defn}

\noindent
A weak automaton is thus a B\"uchi automaton such that each of the
strongly connected components of its graph contains either only
accepting or only non-accepting states.\\
\newline
Not all $\omega$-regular languages can be accepted by deterministic
weak B\"uchi automata, not even by nondeterministic weak
automata. However, there are algorithmic advantages to working with
weak automata : deterministic weak automata can be complemented simply
by inverting their accepting and non-accepting states; and there
exists a simple determinization procedure for weak
automata~\cite{Saf92}, which produces B\"uchi automata that are
deterministic, but generally not weak. Nevertheless, if the
represented language can be accepted by a deterministic weak
automaton, the result of the determinization procedure will be {\em
  inherently weak\/} according to the definition below~\cite{BJW01}
and thus easily transformed into a weak automaton.

\begin{defn}
A B\"uchi automaton is {\em inherently weak\/} if none of the
reachable strongly connected components of its transition graph
contain both accepting (visiting at least one accepting state) and
non-accepting (not visiting any accepting state) cycles.
\end{defn}

This gives us a pragmatic way of staying within the realm of
deterministic weak B\"uchi automata. We start with sets represented by
such automata. This is preserved by union, intersection, synchronous
product, and complementation operations. If a projection is needed,
the result is determinized by the known simple procedure. Then, either
the result is inherently weak and we can proceed, or it is not and we
are forced to use the classical algorithms for B\"uchi automata. The
latter cases might never occur, for instance if we are working with
automata representing sets of reals definable in the first-order
theory of linear constraints~\cite{BJW01}.\\
\newline
A final advantage of weak deterministic B\"uchi automata is that they
admit a minimal form, which is unique up to isomorphism~\cite{Loe01}.

\subsection{Transducers}

In this paper, we will consider relations that are defined over sets
of words. We use the following definitions taken from
\cite{Nil01}. For a finite-word (respectively, infinite-word) language
$L$ over $\Sigma^n$, we denote by ${\lfloor}L{\rfloor}$ the
finite-word (respectively, infinite-word) relation over $\Sigma^n$
consisting of the set of tuples $(w_1,w_2,{\dots},w_n)$ such that
$w_1\bar{{\times}}w_2\bar{{\times}}\dots\bar{{\times}}w_n$ is in
$L$. The arity of such a relation is $n$. Note that for $n=1$, we have
that $L={\lfloor}L{\rfloor}$. The relation $R_{id}$ is the {\em
  identity relation}, i.e.,
$R_{id}={\lbrace}(w_1,w_2,{\dots},w_n)|w_1=w_2={\dots}=w_n{\rbrace}$. A
relation $R$ defined over $\Sigma^n$ is \\{\em ($\omega$-)regular} if
there exists a ($\omega$-)regular language $L$ over $\Sigma^n$ such
that ${\lfloor}L{\rfloor}=R$.\\
\newline
We now introduce transducers that are automata for
representing ($\omega$-)regular relations over $\Sigma^2$.

\begin{defn}
\label{definition-transdu}
A transducer over $\Sigma^2$ is an automaton $T$ over $\Sigma^2$ given
by $(Q,\Sigma^2,$\\$ Q_0,\triangle, F)$, where
\begin{itemize}
\item
$Q$ is the {\em finite} set of {\em states\/},
\item
$\Sigma^2$ is the {\em finite} alphabet,
\item
$Q_0\,\subseteq\,Q$ is the set of {\em initial states\/},
\item
$\triangle: Q\times \Sigma^2\times
Q$ is the {\em transition relation\/}, and
\item
$F\,\subseteq\,Q$ is the set of accepting states (the states that are not
in $F$ are the {\em nonaccepting} states).
\end{itemize}
\end{defn}

\noindent
Given an alphabet $\Sigma$, the transducer representing the identity
relation over $\Sigma^2$ is denoted $T_{id}^{\Sigma}$ (or $T_{id}$
when $\Sigma$ is clear from the context). All the concepts and
operations defined for finite automata can be used with
transducers. The only reason to particularize this class of automata
is that some operations, such as composition, are specific to
relations. In the sequel, we use the term ``transducer'' instead of
``automaton'' when using the automaton as a representation of a
relation rather than as a representation of a language. We sometimes
abuse the notations and write $(w_1,w_2)\in T$ instead of
$(w_1,w_2)\in {\lfloor}L(T){\rfloor}$. Given a pair $(w_1,w_2)\in T$,
$w_1$ is the {\em input word}, and $w_2$ is the {\em output word}. The
transducers we consider here are often called {\em
  structure-preserving}, which means that when following a transition,
a symbol of the input word is replaced by exactly one symbol of the
output word.\\
\newline
Given two transducers $T_1$ and $T_2$ over the alphabet $\Sigma$ that
represents two relations $R_1$ and $R_2$, respectively. The {\em
  composition} of $T_1$ by $T_2$, denoted $T_2\circ T_1$ is the
transducer that represents the relation $R_2\circ R_1$. We denote by
$T_1^i$ ($i\in \nats_0$) the transducer that represents the relation
$R_1^i$. The {\em transitive closure} of $T$ is
$T^+=\bigcup_{i=1}^{\infty}T^i$; its {\em reflexive transitive
  closure} is $T^*=T^+\cup T_{id}$. The transducer $T$ is {\em
  reflexive} if and only if $L(T_{id})\,\subseteq\,L(T)$. Given an
automaton $A$ over $\Sigma$ that represents a set $S$, we denote by
$T(A)$ the automaton representing the {\em image} of $A$ by $T$, i.e.,
an automaton for the set $R(S)$.\\
\newline
Let $T_1$ and $T_2$ be two finite-word (respectively, B\"uchi)
transducers defined over $\Sigma^2$ and let $A$ be a finite-word
automaton (respectively, B\"uchi) automaton defined over $\Sigma$. We
observe that $T_2\circ T_1 = \pi_{\not= 2}{\lbrack}(T_1 \bar{{\times}}
T_{id}^\Sigma)\cap (T_{id}^\Sigma \bar{{\times}} T_2){\rbrack}$ and
$T(A)=\pi_{\not= 1}{\lbrack}(A^{\Sigma} \bar{{\times}} \Sigma)\cap
T{\rbrack}$, where $A^{\Sigma}$ is an automaton accepting $\Sigma^*$
(respectively, $\Sigma^{\omega}$). As a consequence, the composition
of two finite-word ((weak) B\"uchi) transducers is a finite-word
transducer. However, the composition of two deterministic weak B\"uchi
transducer is a weak B\"uchi transducer whose deterministic version
may not be weak. A same observation can be made about the composition
of a transducer with an automaton.

\section{Systems models and ($\omega$)-Regular Model Checking}
\label{paper-model}

\subsection{The Framework}

In this section, we recall the definition of state-transition system,
that is the abstraction formalism which is generally used to describe
programs. We then present an automata-based encoding of
state-transition systems. Finally, the properties of this encoding are
discussed.

\subsubsection{State-transition Systems}
\label{mod-sec}
Systems are often modeled as {\em state-transition
systems}.

\begin{defn}
A state-transition system is a tuple $(S,S_0,R)$, where
\begin{itemize}
\itemsep0cm
\item
$S$ is a (possibly infinite) set of {\em states},
\item
$S_0\,\subseteq\,S$ is a (possibly infinite) set of {\em initial
states}, and
\item
$R\,{\subseteq}\,S\times S$ is a (possibly infinite) {\em
reachability relation} that describes the transitions between the
states of the system.
\end{itemize}
\end{defn}

\noindent
Let $\TS=(S,S_0,R)$ be a state-transition system. If $(s,s')\in R$,
then we say that there is a {\em transition} from $s$ (the {\em
  origin}) to $s'$ (the {\em destination}). Given two states $s,s'\in
S$, we write $s{\rightarrow}_R~s'$ if and only if $(s,s')\in R$. A
state $s'\in S$ is said to be {\em reachable} from a state $s\in S$ if
there exists $k>0$ and states $s_0,s_1,s_2,{\dots},s_{k-1}\in S$ such
that $s_0=s$, $s_{k-1}=s'$ and $s_i{\rightarrow}_R s_{i+1}$, for all
$0\,{\leq}\,i<k-1$. The fact that $(s,s')$ belongs to the reflexive
transitive closure $R^*$ of $R$ is denoted by
$s{\rightarrow}_{R}^*~s'$. A state $s\in S$ is {\em reachable} if it
is reachable from a state in $S_0$. The set of all reachable states of
$\TS$ is denoted $S_R^{\TS}$. The {\em state space}
$(S_R^{\TS},R_R^{\TS})$ of $\TS$ is the (possibly infinite) graph
whose nodes are the reachable states of $\TS$, and whose edges
$R_R^\TS$ are given by $R\cap (S_R^{\TS}\times S_R^{\TS})$.  We say
that $\TS$ is {\em finite} if $S_R^{\TS}$ is finite, it is {\em
  infinite} otherwise. $\TS$ is said to be {\em locally-finite} if and
only if any executions from any state in $S_R^{\TS}$ can only goes
thought a finite number of distinct states. A finite {\em execution}
$\pi$ of $\TS$ is a mapping $\pi:
{\lbrace}0,\dots,n-1{\rbrace}\rightarrow S$ such that $\pi(0)\in S_0$
and for all $0\,{\leq}\,i<n-1$, $\pi(i)\rightarrow_R \pi(i+1)$. A
finite execution is often represented by
$\pi=\pi(0)\pi(1)\pi(2){\dots}\pi(n-1)$. An infinite execution $\pi$
of $\TS$ is a mapping $\pi: \nats\rightarrow S$ such that $\pi(0)\in
S_0$ and for all $i\,{\geq}\,0$, $\pi(i)\rightarrow_R \pi(i+1)$. An
infinite execution is often represented by
$\pi=\pi(0)\pi(1)\pi(2)\dots$. In the rest of this paper, we consider
systems whose executions are {\em all} infinite.\\
\newline One distinguishes between two types of properties.

\begin{enumerate}
\item
{\em Reachability properties.} We assume that a reachability property
$\varphi$ is described as a set of states
$S_{\varphi}\,\subseteq\,S$. The system $\TS$ satisfies $\varphi$ if and only if
$S_R^{\TS}\,\subseteq\,S_{\varphi}$. Verifying reachability properties
thus reduces to computing the set of reachable states.
\item {\em Linear temporal properties.} We assume that a linear
  temporal property $\varphi$ is described as a set of executions
  $\pi_{\varphi}$, which are often represented by a B\"uchi
  automaton. The system $\TS$ satisfies $\varphi$ if and only if each
  of its executions belongs to $\pi_{\varphi}$. In general, the
  verification of linear temporal properties does not reduce to the
  computation of the set of reachable states of the system.
\end{enumerate}

\subsubsection{($\omega$)-Regular Model Checking}
\label{models-regular}

In this paper, we suppose that states of state-transition systems are
encoded by words over a fixed alphabet. If the states are encoded by
finite words, then sets of states can be represented by finite-word
automata and relations between states by finite-word transducers. This
setting is referred to as {\em regular model
  checking}\,\cite{KMMPS97,WB98}. If the states are encoded by
infinite words, then sets of states can be represented by
deterministic weak B\"uchi automata and relations between states by
deterministic weak B\"uchi transducers. This setting is referred to as
{\em $\omega$-regular model checking}\,\cite{BLW04a}. Formally, a
finite automata-based representation of a state-transition system can
be defined as follows.\\
\newline
\begin{defn}
A {\em ($\omega$-)regular system} for a state-transition system
$\TS=(S,S_0,R)$ is a triple $M=(\Sigma, A_{S_0}, T_R)$, where
\begin{itemize}
\itemsep0cm
\item
$\Sigma$ is a finite alphabet over which the states are encoded as
finite (respectively infinite) words;
\item
$A_{S_0}$ is a deterministic finite-word (respectively deterministic weak
  B\"uchi) automaton over $\Sigma$ that represents $S_0$;
\item
$T_R$ is a deterministic finite-word (respectively deterministic weak
  B\"uchi) transducer over $\Sigma^2$ that represents $R$.
\end{itemize}
\end{defn}

\noindent
States being represented by words, the notion of set of states,
initial states, reachability relation, computation, reachable state,
locally-finite for ($\omega$-)regular systems are defined identically
to those of the corresponding state-transition system. There are many
state-transition systems whose sets of states cannot be encoded by
($\omega$)-regular languages{\footnote{Indeed, there are uncountably
many subsets of an infinite set of states, but only countably many
finite strings of bits.}}. Consequently, there are many
state-transition systems for which there exists no corresponding
($\omega$-)regular system.

In the finite-word case, an execution of the system is an infinite
sequence of same-length finite words over $\Sigma$. The regular model
checking framework was first used to represent parametric systems
\cite{AJMd02,KMMPS97,ABJN99}. The framework can also be used to
represent various other models, which includes linear integer
systems\,\cite{WB95,WB00}, FIFO-queues systems\,\cite{BG96}, {\em XML}
specifications\,\cite{BHRV06b,Td06}, and heap
analysis\,\cite{BHMV05,BHRV06b}.\\
\newline We now give insight about how to represent
parametric systems. Let $P$ be a process represented by a finite
state-transition system. A parametric system for $P$ is an infinite
family $S={\lbrace}S_n{\rbrace}_{n=0}^{\infty}$ of networks where for
a fixed $n$, $S_n$ is an {\em instance} of $S$, {\emph{i.e.} a network
  composed of $n$ copies of $P$ that work together in parallel. In the
  regular model checking framework, the finite set of states of each
  process is given as an alphabet $\Sigma$. Each state of an instance
  of the system can then be encoded as a finite word
  $w=w(0){\dots}w(n-1)$ over $\Sigma$, where $w(i-1)$ encodes the
  current state of the $i$th copy of $P$. Sets of states of several
  instances can thus be encoded together by finite-word
  automata. Observe that the states of an instance $S_n$ are all
  encoded with words of the same length. Consequently, relations
  between states in $S_n$ can be represented by binary finite-word
  relations, and eventually by transducers.


\begin{exmp}
\label{example-token}
Consider a simple example of parametric network of identical processes
implementing a token ring algorithm. Each of these processes can be
either in idle or in critical mode, depending on whether or not it
owns the unique token. Two neighboring processes can communicate with
each other as follows: a process owning the token can give it to its
right-hand neighbor. We consider the alphabet
$\Sigma={\lbrace}N,T{\rbrace}$. Each process can be in one of the two
following states : $T$ (has the token) or $N$ (does not have the
token). Given a word $w\in \Sigma^*$ with $|w|=n$ (meaning that $n$
processes are involved in the execution), we assume that the process
whose states are encoded in position $w(0)$ is the right-hand neighbor
of the one whose states are encoded in position $w(n-1)$. The
transition relation can be encoded as the union of two {\em
  reachability} relations that are the following:
\begin{itemize}
\item
$(N,N)^*(T,N)(N,T)(N,N)^*$ to describe the move of the
token from $w(0)$ to $w(n-1)$, and
\item
$(N,T)(N,N)^*(T,N)$ to describe the move of the token from
$w(n-1)$ to $w(0)$.
\end{itemize}

The set of all possible initial states where the first process has the
token is described by $TN^*$. 
\end{exmp}

In the infinite-word case, an execution of the system is an infinite
sequence of infinite words over $\Sigma$. The $\omega$-regular model
checking framework has been used for handling systems with both
integer and real variables~\cite{BW02,BJW05}, such as linear hybrid
systems with a constant derivative (see examples in \cite{ACHH95} or
in \cite{BLW04b,Leg07}).\\
\newline
Verifying reachability properties of state-transition systems using
their ($\omega$-)regular representation can easily be conducted with
simple automata-based manipulations, assuming the existence of
finite-word (respectively weak B\"uchi) automata for representing both the
set of reachable states and the property. Computing an automaton that
represents the set of reachable states can be reduced to the {\em
  ($\omega$-)regular reachability problems} defined hereafter.

\begin{defn}  
Let $A$ be a deterministic finite-word (respectively weak B\"uchi) automaton,
and $T$ be a deterministic finite-word (respectively weak B\"uchi)
transducer. The ($\omega$-)regular reachability problems for $A$ and
$T$ are the following:
\begin{enumerate}
\item
{\em Computing $T^*(A)$:} the goal is to compute a finite-word
(respectively weak B\"uchi) automaton representing $T^*(A)$. If $A$
represents a set of states $S$ and $T$ a relation $R$, then $T^*(A)$
represents the set of states that can be reached from $S$ by applying
$R$ an arbitrary number of times;
\item {\em Computing $T^*$:} the goal is to compute a finite-word
  (respectively weak B\"uchi) transducer representing the reflexive
  transitive closure of $T$. If $T$ represents a reachability relation
  $R$, then $T^*$ represents its closure $R^*$.
\end{enumerate}
\end{defn}

Being able to compute $T^*(A)$ is clearly enough for verifying
reachability properties. On the other hand, we will see that the
computation of $T^*$ is generally incontrovertible when considering
the verification of temporal properties. In the rest of this paper, we
propose techniques that reduce the verification of several classes of
linear temporal properties to the resolution of the ($\omega$-)regular
reachability problems over an augmented system.

\subsection{On Solving ($\omega$-)Regular Reachability Problems}

Among the techniques to compute $T^*(A)$ and $T^*$, one distinguishes
between {\em domain specific and generic} techniques. Domain specific
techniques exploit the specific properties and representations of the
domain being considered and were for instance obtained for systems
with FIFO-queues in~\cite{BG96,BH97}, for systems with integers and
reals in~\cite{Boi98,BW02,BHJ03}, for pushdown systems
in~\cite{FWW97,BEM97}, and for lossy queues in~\cite{AJ96}.  Generic
techniques consider automata-based representations and provide
algorithms that operate directly on these representations, mostly
disregarding the domain for which it is used. There are various
generic techniques to computing $T^*(A)$ and $T^*$ when considering
$T$ and $A$ to be finite-word automata
(e.g. \cite{BJNT00,DLS02,BLW03}).  The
$\omega$-regular reachability problems can be addressed with the
technique introduced in \cite{BLW04a}.



\subsection{Convention, Concepts, and Observations}
\label{liveness-observations}

This section introduces some concepts and observations that will be
used throughout the rest of the paper. We first introduce {\em B\"uchi
  ($\omega$-)regular systems}.

\begin{defn}
A B\"uchi ($\omega$-)regular system is a tuple $(M,F)$, where
$M=(\Sigma, A_{S_0},$\\$ T_R)$ is a ($\omega$-)regular system, and $F$
is a deterministic finite-word (resp. deterministic weak B\"uchi)
automaton called the {\em B\"uchi acceptance condition}.
\end{defn}

\noindent
The notions of set of states, initial states, reachability relation,
computation, reachable state, and locally-finite for B\"uchi
($\omega$-)regular system $(M,F)$ are defined exactly as those of its
underlying ($\omega$-)regular system $M=(\Sigma, A_{S_0}, T_R)$. An
infinite computation $\pi=\pi(0)\pi(1)\dots$ of $(M,F)$ is {\em
  accepting} if and only if there are infinitely many $i$ such that
$\pi(i)\in L(F)$. We say that $(M,F)$ is {\em empty} if all its
infinite executions are non-accepting. In the rest of the paper, we
abuse the notations and write $(\Sigma, A_{S_0}, T_R,F)$ instead of
$(M,F)$.\\
\newline
We now reason on infinite executions. Consider a ($\omega$-)regular
system $M=(\Sigma, A_{S_0}, T_R)$ that encodes a state-transition
system $\TS=(S,S_0,R)$. The fact that $T_R$ is structure-preserving
does not imply that $M$ is locally-finite. Indeed, as it is
illustrated with the following example, each state of $\TS$ can
potentially be associated to an infinite set of encodings.

\begin{exmp}
Following the framework of \cite{WB00}, the digit $5$ can be encoded
in base $2$ as $0101$, or as $00101$, or as $000101$, ..., and in fact
by any word in the set $0^+101$.
\end{exmp}

By definition, parametric systems are always locally-finite. Indeed,
the number of finite-state processes is fixed during the whole
execution. This makes it impossible to visit an infinite number of different
states. Most other classes of infinite-state systems can either be
locally-finite or not, depending on their specifications.

\begin{exmp}
An integer system that continuously adds $1$ to a variable $x$ up to a
constant value is locally-finite. However, if there is no bound on the
value of $x$, then the system is not locally-finite.
\end{exmp}

\noindent
Unfortunately, testing whether a system is locally-finite is an
undecidable problem. As a consequence only partial solutions can be
proposed. In the rest of this section, we propose such a solution that
is based on a reduction to the ($\omega$)-regular reachability
problems over an augmented system. Our solution is formalized with the
following theorem.

\begin{thm}
\label{other-precis-states}
Consider a state-transition system $\TS=(S,S_0,R)$ and the following
sets
\begin{itemize}
\item
$S_0^a=S_0\times {\lbrace}0{\rbrace}$, 
\item $R^a={\lbrace}((s,i),(s',i+1))\mid (i\in \nats)((s,s')\in
  R\setminus R_{id}){\rbrace}$,
\item
$S_{lf}={\lbrace}s\in S\mid \exists i,\forall (j>i),\neg \exists s' ((s,0),(s',j))\in (R^a)^*{\rbrace}$.
\end{itemize}
If $S_0^a\,\subseteq\,S_{lf}$, then $\TS$ is locally-finite.
\end{thm} 

\begin{pf} 
Direct by construction.
\end{pf}

\noindent
The procedure sketched above requires to compute the set $S_{lf}$. In
the ($\omega$-)regular model checking framework, this computation can
easily be performed when both $(R^a)^*$ and $S_0^a$ represent
solutions of Presburger arithmetic formulas\,\cite{WB00,BJW05}.


\section{Linear Temporal Properties in Regular Model Checking}
\label{temporal-rmc1}
\subsection{Definitions}
\label{temporal-rmc11}

In this section we propose a methodology to verify linear temporal
properties of state-transition systems that are represented in the
regular model checking framework. Our first step is a symbolic
representation for linear temporal properties in this framework. We
propose the following definitions.

\begin{defn}
Given an alphabet $\Sigma$, a {\em state property} is a set
$\cop \subseteq \Sigma^*$ that can be represented by a finite-word
automaton.
\end{defn}

\begin{defn}
Let $\it{COP}$ be a finite set of state properties. A {\em global
system property} over $\it{COP}$ is a set $\gsp \subseteq
(2^{\it{COP}})^\omega$, i.e. a set of infinite sequences of
state properties, that can be represented by a B\"uchi automaton.
\end{defn}

\noindent
Assume a set of state properties $\it{COP}$, and a global system
property $\gsp$ defined over $\it{COP}$. An execution
$\pi=w_0w_1w_2w_3w_4\dots$ of a regular system $M$ satisfies $\gsp$,
denoted $\pi \models \gsp$, if and only if $\fcop(w_0)\fcop(w_1)\cdots
\in \gsp$, where $\fcop(w)=\{\cop_i\in \it{COP} \mid w \models
\cop_i\}$. We say that $M$ satisfies $\gsp$, denoted $M\models \gsp$,
if and only if all its executions satisfy the property.\\
\newline
\noindent
The definition of global system properties is illustrated in Figure
\ref{global-figure}.

\begin{figure}
\begin{center}
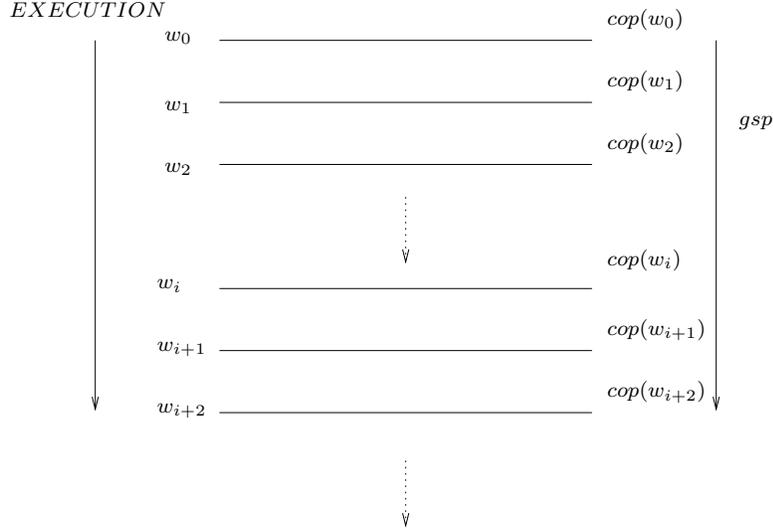
\caption{Global system properties: an illustration.}
\label{global-figure}
\end{center}
\end{figure}

\begin{rem}
\label{rem-global1}
Any {\em Linear Temporal Logic property}\footnote{We assume the reader
  is familiar with the syntax, the semantic, and the notations of the
  linear temporal logic introduced in \cite{Pnu77}. We recall the
  shortcuts for the temporal operators that are $\Box$ for ``always'',
  $\Diamond$ for eventually, and $\bigcirc$ for ``next''.}  (LTL in
short) whose atomic propositions are represented by sets of states is
thus a global system property. The set of LTL properties whose atomic
propositions are represented by sets of states is a strict subset of
the set of global system properties.
\end{rem}

\subsection{Verification}
\label{model-global-regular}

Assume a regular system $M=(\Sigma, A_{S_0}, T_R)$, a set of state
properties $\it{COP}=\{\cop_1,\ldots\cop_k\}$, and a global system
property $\gsp$ defined over $\it{COP}$. Suppose that each $\cop_i\in
\it{COP}$ is represented by a complete deterministic finite-word
automaton $A_{\cop_i}=( Q_{\cop_i}, \Sigma,
{q_0}_{\cop_i},\delta_{\cop_i},F_{\cop_i})$. We extend the automata
theoretic approach of \cite{VW86} towards a semi-algorithm to test
whether $M$ satisfies $\gsp$. Our approach consists in three
successive steps that are the following:

\begin{enumerate}
\item
Computing a complete B\"uchi automaton $A_\ogsp=(Q_\ogsp,
2^{\it{COP}}, {q_0}_\ogsp,$\\$\triangle_\ogsp,F_\ogsp)$ representing
the negation of the property $\gsp$, {\emph{i.e.}}\\
$(2^{\it{COP}})^\omega\setminus \gsp$;
\item
Building a B\"uchi regular system $M^a_\ogsp = (\Sigma^a, A^a_{S_0},
T^a_{R},F^a)$ whose accepting executions correspond to those of $M$
that are accepted by $A_\ogsp$;
\item
Testing whether $M^a_\ogsp$ is empty or not. By construction, $M$ satisfies
$\gsp$ if and only if $M^a_\ogsp$ is empty.
\end{enumerate}

The property $\gsp$ being (by definition) representable by a B\"uchi
automaton, one can always compute the automaton $A_\ogsp$. We now
focus on the two other problems. The system $M^a_\ogsp$ can be built
by taking the product between the states of $M$ and those of
$A_\ogsp$. Given $w,w'\in \Sigma^*$ and $q_\ogsp,q_\ogsp'\in Q_\ogsp$,
the product must ensure that one can move from the pair $(w,q_\ogsp)$
to the pair $(w',q_\ogsp')$ if and only if (1) $(w,w')\in T_R$, and
(2) $(q_\ogsp,\fcop(w),q_\ogsp')\in \triangle_\ogsp$. Since the set of
states of $M$ may be infinite, we have to work with a symbolic
representation of $\fcop$. We propose to represent $\fcop$ implicitly
by associating to each pair $(w,q)$ the set $\it{COP}_i$ such that
$\fcop(w)=\it{COP}_i$. Hence a state of the product is now a triple
$(w,q_\ogsp,\it{COP}_i)$ such that (1) $w\in \Sigma^*$, (2)
$q_\ogsp\in Q_\ogsp $, and (3) $\fcop(w)=\it{COP}_i$. Each triple
$(w,q_\ogsp,\it{COP}_i)$ has to be encoded by a finite word over an
extended alphabet. The solution is to label the last symbol of $w$
with $\it{COP}_i$ and $q_\ogsp$, and the other symbols by
$\bot$. Hence, we define the augmented alphabet to be

$$\Sigma^a = \Sigma \times ( Q_\ogsp \cup \{\bot\})\times (2^{\it{COP}}
\cup \{\bot\}).$$

\noindent
Given a word $w^a\in (\Sigma^a)^*$, we denote by $\Pi_\Sigma(w^a)$,
the word $w\in \Sigma^*$ obtained from $w^a$ by removing all the
symbols that do not belong to $\Sigma$. As an example, given
$w^a=(w(0),\bot,\bot)(w(1),\bot,\bot)\cdots(w(n-1),q,\lambda)$ with $q\in
Q_\ogsp$, $\lambda\in 2^{COP},\Pi_\Sigma(w^a)=w(0)w(1)\cdots
w(n-1)$.\\
\newline
\noindent
An execution $\pi^a=w^a_0w^a_1w^a_2\dots$ of $M^a_\ogsp$ is an
infinite sequence of finite words over $\Sigma^a$. This sequence has
to satisfy the following four requirements:

\begin{enumerate}
\item
For each $i\,{\geq}\,1$ $(\Pi_\Sigma(w^a_{i-1}),\Pi_\Sigma(w^a_i))\in
T_R$, which ensures that the transitions of $M^a_\ogsp$ are compatible
with the transition relation of $M$;
\item
For each $i\,{\geq}\,0$, $w_i^a\in (\Sigma\times \bot\times
\bot)^*(\Sigma\times Q_\ogsp\times 2^{\it{COP}})$;
\item
For each $i\,{\geq}\,0$ and
$w_i^a=(w_i(0),\bot,\bot)(w_i(1),\bot,\bot)\cdots(w_i(n-1),q_{i\ogsp},\it{COP}_i)$,
$\fcop(\Pi_\Sigma(w_i^a))=\it{COP}_i$;
\item For each $i\,{\geq}\,1$ and
  $w_{i-1}^a=(w_{i-1}(0),\bot,\bot)(w_{i-1}(1),\bot,\bot)\cdots(w_{i-1}(n-1),q_{i-1\ogsp},\it{COP}_{i-1})$,
  and
  $w_i^a=(w_i(0),\bot,\bot)(w_i(1),\bot,\bot)\cdots(w_i(n-1),$\\$q_{i\ogsp},\it{COP}_i)$,
  we have $(q_{i-1\ogsp},\it{COP}_{i-1},q_{i\ogsp})\in
  \triangle_\ogsp$, this to ensure that the infinite sequence of
  labellings from $2^{\it{COP}}$ and $Q_\ogsp$ form a run of the
  automaton $A_\ogsp$.
\end{enumerate}

\noindent
We have to build automata for $A^a_{S_0}$, $F^a$, and $T_{R}^a$ in
such a way that the four requirements above are satisfied.

Let $T_R=(Q_R,\Sigma^2, {q_{0R}}, \delta_R, F_R)$, the transducer
$T^a_{R}=(Q_R^a,(\Sigma^a)^2,q_{0R}^a,\triangle_R^a,F_R^a)$ is built
as follows:

\begin{itemize}
\item
The set of states $Q_{R}^a$ is $Q_{R}^a = Q_R \times \prod_{1\leq i
\leq k} Q_{\cop_i}\times \boolset,$ the last Boolean being used to
remember if non $\bot$ labellings have been seen and $\prod_{1\leq i
\leq k} Q_{\cop_i}$ is used to run the automata representing the
state properties, this to ensure that each state of $M^a_\ogsp$ is
associated to the set of state properties it satisfies;
\item
The initial state is
$q_{0R}^a=({q_{0R}},{q_0}_{\cop_1},\ldots, {q_0}_{\cop_k}, 0)$;
\item
The transition relation $\triangle_R^a$ is defined by
$$({q}_{R}',{q}_{\cop_1}',\ldots, {q}_{\cop_k}',b') \in
\triangle_{R}^a(({q}_{R},{q}_{\cop_1},\ldots, {q}_{\cop_k},b),
((a_1,\alpha_1,\lambda_1), (a_2,\alpha_2,\lambda_2)))$$ if and only if
\begin{itemize}
\item
${q}_{R}' \in \delta_R(q_R,(a_1,a_2))$ and $q_{\cop_i}' =
\delta_{\cop_i}(q_{\cop_i},a_1)$, for $1\leq i \leq k$,
\item
$b'=1$ if and only if $\lambda_1$, $\alpha_1$, $\lambda_2$, and
$\alpha_2$ are not equal to $\bot$ and, in this case, $\alpha_2 \in
\triangle_\ogsp(\alpha_1,\lambda_1),$ which checks that we have a run
of $A_\ogsp$ and, for $1\leq i \leq k$, $q_{cop_i}' \in F_{cop_i}\ \
\mbox{if and only if}\ \ cop_i \in \lambda_1,$ which checks that the
label $\lambda_1$ matches the result of running the automata
$A_{\cop_i}$ on the state (this justify the need for each $A_{\cop_i}$
to be deterministic and complete);
\end{itemize}
\item
The set of accepting states $F_R^a$ is defined as $F_R \times
\prod_{1\leq i \leq k} Q_{\cop_i}\times \{1\}.$
\end{itemize}

\noindent
The definition of $T^a_R$ ensures that requirements (1) (3) and (4)
are satisfied.\\
\newline
The set of initial states of $M^a_\ogsp$ contains states of the following form:
$$(w(0),\bot,\bot) \cdots (w(n-2),\bot,\bot)(w(n-1), {q_0}_\ogsp,
\lambda),$$ where $w(0)\cdots w(n-1) \in L(A_{S_0})$ and $\lambda$ is any
element of $2^{\it{COP}}$. This definition combined with the one of
$T^a_R$ ensures that the second requirement on the executions of
$M^a_\ogsp$ is always satisfied. The set of initial states can be
represented by a finite-word automaton $A_{S_0}^a$ that is given by
$A_{S_0}{\bar{\times}}A_\bot$, where $A_\bot$ is the automaton
representing the set $(\bot\times \bot)^*({q_0}_\ogsp\times
2^{\it{COP}})$.\\
\newline
The set of accepting states of $M^a_\ogsp$ is defined as follows:
$$(\Sigma\times \{\bot\}{\times}\{\bot\})^*(\Sigma\times F_\ogsp
\times 2^{\it{COP}}).$$ We directly see that this set can be
represented by a finite-word automaton $F^a$.

\begin{thm}
The B\"uchi regular system $M^a_\ogsp$ has an accepting execution of
the form $\pi^a=\pi^a(0)\pi^a(1)\dots$ if and only if the execution
$\pi=w_0w_1w_2\dots$ of $M$, where $w_i=\Pi_\Sigma(\pi^a(i))$
($\forall i$), does not satisfy $\gsp$.
\end{thm}

\begin{pf}
Follows from the construction above.
\end{pf}

The next step is to test whether $M^a_\ogsp$ is empty or not. If $M$
is locally-finite, then $M^a_\ogsp$ is also locally-finite and
checking the emptiness of $M^a_\ogsp$ can be reduced to solving the
regular reachability problems. We have the following result.

\begin{prop}
\label{prop-proof}
If $M^a_\ogsp$ is locally finite, then it is empty if and only if
$$L((T^a_R)^*(A^a_{S_0}) \cap F^a \cap \Pi_{\not=2}((T^a_R)^+ \cap
T_{id}))=\emptyset.$$
\end{prop}

\begin{pf}
  Directe by observing that since $M^a_\ogsp$ is locally-finite, any
  of its accepting execution must repeatly reach a given state in
  $F^a$.
\end{pf}

If $M^a_\ogsp$ is not locally-finite, then we cannot reduce the
problem of deciding if it has an infinite accepting execution to the
one of finding reachable accepting loops. Indeed, in this case, an
infinite execution could never visit the same state twice. Therefore,
our approach is to search for a reachable state $w$ from which it is
possible to nontrivially reach some state $w'$ such that (1) the path
from $w$ to $w'$ visits a repeating state of $A_{\neg gsp}$, and (2)
$w'$ has at least the same execution paths as $w$. To check the
condition (2), we check actually for a stronger condition which is the
fact that $w'$ must {\em simulate} $w$.\\
\newline
We define the {\em greatest simulation relation} over $M^a_\ogsp$
which is compatible with the set of state properties $\it{COP}$ to be the
relation $Sim$ defined as the limit of the (possibly infinite)
decreasing sequence of relations $Sim_0,Sim_1, Sim_2,\dots$ with

\begin{eqnarray}
Sim_0 & = & \{(w_1^a,w_2^a)\} \mid w_1^a{\bar{\times}}w_2^a\in (\Sigma^a\times
\Sigma^a)^* \ \wedge \fcop(\Pi_\Sigma(w_1^a))=\fcop(\Pi_\Sigma(w_2^a))\\
Sim_{k+1} & = & Sim_k \cap \{ (w_1^a,w_2^a) \in Sim_k \; \mid \; \\
&&\ \ \ \forall w_3^a .((w_1^a,w_3^a) \in T^a_R 
\Rightarrow \exists w_4^a .(w_2^a,w_4^a) \in T^a_R 
\wedge (w_3^a,w_4^a) \in Sim_k )\} \nonumber, \forall k\in \nats
\end{eqnarray}

\noindent
The {\em complement} of $Sim$, denoted $\neg Sim$, is the set
${\lbrace}(w_1^a,w_2^a)\mid (w_1^a{\bar{\times}}w_2^a \in
(\Sigma^a\times \Sigma^a)^*)\,{\wedge}$\\$\,((w_1^a,w_2^a)\notin
Sim)){\rbrace}$. The {\em greatest simulation equivalence} over
$M^a_\ogsp$ which is compatible with $\it{COP}$ is the relation
$\widetilde{Sim} = Sim \cap Sim^{-1}$. Observe that $Sim_0$ can be
represented by a finite-word transducer over the alphabet
$(\Sigma^a)^2$. Since, for each $k\,{\geq}\,0$, the relation
$Sim_{k+1}$ is defined in terms of the relations
${\lbrack}L(T_R^a){\rbrack}$ and $Sim_k$ using Boolean operations and
projections (needed to apply the quantifiers), it can be represented
by a finite-word transducer over the alphabet
$(\Sigma^a)^2$. Moreover, if $Sim_k$ can be represented by a
transducer, then its complement and inverse can also be represented in
the same way.\\
\newline
Assume that $Sim$ and $(\Sigma^a)^*$ are respectively represented by a
transducer $T_{Sim}$ and an automaton $A^{(\Sigma^a)^*}$. We have the
following result.

\begin{prop}
\label{simu-prop1}
If
\begin{equation*}
L((T^a_R)^*(A^a_{S_0}) \cap \Pi_{\not=2}((T^a_R)^+ \cap (A^{(\Sigma^a)^*}{\bar{\times}} F^a) \cap T_{Sim}))\not= \emptyset,
\end{equation*}
then $M^a_\ogsp$ has an infinite execution that does not satisfy $\gsp$.
\end{prop}

\begin{pf}
The set $L(\Pi_{\not=2}((T^a_R)^+ \cap (A^{(\Sigma^a)^*}{\bar{\times}}
F^a) \cap T_{Sim}))$ is the set of states $w$ from which it is
possible to reach an accepting state $w'$ such that $w'$ simulates
$w$. Since $w'$ simulates $w$, one can reach from $w'$ an other
accepting state $w''$ that simulates $w'$ and, inductively, there
exists an execution that infinitely often goes through an accepting
state.
\end{pf}



The main issue is now to determine whether the iterative computation
of $Sim$ terminates and can be represented by an automaton. We
consider the two following cases.

\subsubsection{Exact Analysis}

We say that $M^a_\ogsp$ has a {\em finite-index simulation} if the
simulation equivalence $\widetilde{Sim}$ has a finite number of
equivalence classes. The following lemma is quite straightforward.

\begin{lem}
The iterative computation of the simulation relation $Sim$ terminates
if and only if $M^a_\ogsp$ has a finite-index simulation.
\end{lem}

If $M^a_\ogsp$ has a finite-index simulation equivalence, then every
infinite execution of $M^a_\ogsp$ must visit infinitely often some of
the equivalence classes. Therefore, we have the following proposition.

\begin{prop}
Assume that the system $M^a_\ogsp$ has a finite-index
simulation. $M^a_\ogsp$ has an accepting execution if and only if
\begin{equation*}
L((T^a_R)^*(A^a_{S_0}) \cap \Pi_{\not=2}((T^a_R)^+ \cap (A^{(\Sigma^a)^*}{\bar{\times}} F^a) \cap T_{Sim}))\not= \emptyset.
\end{equation*}
\end{prop}

\noindent
However, the system $M^a_\ogsp$ is in general not finite-index
simulation. Moreover this property is undecidable. Therefore, we adopt
an approach based on the use of over/lower approximations of $Sim$.




\subsubsection{Using lower approximations:}

Instead of computing the decreasing sequence of relations $(Sim_i:
i\in \nats)$, we can compute the {\em increasing} sequence of their
negations $(\neg Sim_i:i\in \nats)$. Then, the computed sequence of
relations is actually an increasing sequence of relations $(N_i:i\in
\nats)$ such that for every $i \geq 0$, $N_i = \neg Sim_i$. Since each
$N_i$ can be represented by a transducer, we can use the
extrapolation-based technique of \cite{BLW03,BLW04a,Leg07}. The
technique can compute an automaton that represents an {\em
  extrapolation} $N^{e_*}$ of the limit $\bigcup_{i=0}^{i=+\infty}N_i$
by observing finite prefixes of the sequence $N_0, N_1, N_2, \dots$. A
sufficient criterion to test whether this extrapolation is safe (does
it contain the limit?)  consists in applying one more time the
construction that builds $Sim_{k+1}$ from $Sim_k$ to the complement of
$N^{e_*}$, and then check if the complement of the result we obtain is
included in $N^{e_*}$. We can use the technique of
\cite{BLW03,BLW04a,Leg07} to compute an upper approximation $N^{e_*}$
of the limit of the sequence $(N_i:i\in \nats)$. The negation of
$N^{e_*}$, denoted $\neg N^{e_*}$, is a lower approximation of
$S$. Let $T_{\neg N^{e_*}}$ be the transducer representing $\neg
N^{e_*}$. If the following condition holds

$$L((T^a_R)^*(A^a_{S_0}) \cap \Pi_{\not=2}((T^a_R)^+ \cap
(A^{(\Sigma^a)^*}{\bar{\times}} F^a) \cap T_{\neg N^{e_*}}))\not=
\emptyset,$$

\noindent
then we can deduce that $M^a_\ogsp$ has an infinite accepting
execution, which means that $M^a_\ogsp$ does not satisfy the property
$gsp$.

\section{Linear Temporal Properties in $\omega$-Regular Model  Checking}
\label{last-thesis}

\subsection{Definitions}

We extend the concept of global system properties from regular to
$\omega$-regular systems. For this, we simply encode state-properties
as sets of infinite words rather than sets of finite words. We propose
the following definitions.

\begin{defn}
Given an alphabet $\Sigma$, a {\em $\omega$-state property} is a set
$\cop \subseteq \Sigma^\omega$ that can be represented by a
deterministic weak B\"uchi automaton.
\end{defn}

\noindent
The choice of using deterministic weak automata to represent
$\omega$-state properties is for technical reasons that will be
clarified in the next section.

\begin{defn}
Let $\it{COP}$ be a finite set of $\omega$-state properties defined
over an alphabet $\Sigma$. An {\em $\omega$-global system property}
over $\it{COP}$ is a set $\gsp \subseteq (2^{\it{COP}})^\omega$,
{\emph{i.e.}} a set of infinite sequences of $\omega$-state
properties, that can be represented by a B\"uchi automaton.
\end{defn}

\noindent
Assume a set of $\omega$-state properties $\it{COP}$, and a global
system property $\gsp$ defined over $\it{COP}$. An execution
$\pi=w_0w_1w_2w_3w_4\dots$ of an $\omega$-regular system $M$ satisfies
$\gsp$, denoted $\pi \models \gsp$, if and only if
$\fcop(w_0)\fcop(w_1)\cdots \in \gsp$, where
$\fcop(w)=\{\cop_i\in \it{COP} \mid w \models \cop_i\}$. We say that
$M$ satisfies $\gsp$, denoted $M\models \gsp$, if and only if all its
executions satisfy the property.\\

\subsection{Verification}

Assume an $\omega$-regular system $M=(\Sigma, A_{S_0}, T_R)$, a set of
$\omega$-state properties $\it{COP}=\{\cop_1,\ldots,\cop_k\}$, and an
$\omega$-global system property $\gsp$ defined over
$\it{COP}$. Suppose that the negation of $\gsp$ can be represented by
a B\"uchi automaton $A_\ogsp=(Q_\ogsp, 2^{\it{COP}},
{q_0}_\ogsp,\triangle_\ogsp, F_\ogsp)$, and that each $\cop_i\in
\it{COP}$ can be represented by a complete deterministic weak B\"uchi
automaton $A_{\cop_i}=( Q_{\cop_i}, \Sigma, {q_0}_{\cop_i},
\delta_{\cop_i}, F_{\cop_i})$.

To test whether $M$ satisfies $\gsp$, we proceed as in Section
\ref{model-global-regular} and build a B\"uchi $\omega$-regular system
$M^a_\ogsp=(\Sigma^a,A^a_{S_0},T_R^a,F^a)$ whose executions correspond
to those of $M$ that do not satisfy $\gsp$. We then check whether
$M^a_\ogsp$ is empty or not. We already provided partial solutions to
test whether a B\"uchi regular system is empty or not, and those
solutions directly extend to B\"uchi $\omega$-regular systems. In the
rest of this section, we mainly focus on the construction of
$M^a_\ogsp$.\\
\newline
The main difference between the present case and the one in Section
\ref{model-global-regular} is that since we are working with
infinite-words, we cannot encode the current state of the B\"uchi
automaton $A_{\neg gsp}$ and the current set of $\omega$-state
properties satisfied only in one position of each word of
$M$. Therefore, we include this information everywhere (in each
position) of the word. We must also ensure that this information is
the same for each position (which is needed to be coherent with the
definition of product between $M$ and $A_\ogsp$). We use the following
augmented alphabet:

$$\Sigma^a = \Sigma \times Q_{\neg gsp} \times 2^{\it{COP}}.$$

\noindent
Let $T_R=(Q_R,\Sigma^2, {Q_{0R}}, \delta_R, F_R)$, the possibly
nondeterministic transducer
$T^a_{R}=(Q_R^a,(\Sigma^a)^2,Q_{0R}^a,\triangle_R^a,F_R^a)$ is built
as follows:

\begin{itemize}
\item
The set of states $Q_{R}^a$ is $Q_{R}^a = Q_R \times \prod_{1\leq i
\leq k} Q_{\cop_i}\times Q_{\ogsp} \times 2^{\it{COP}}$. Instead of
the Boolean variable, we have to store the state $A_\ogsp$ and the set
$\it{COP}_i\in 2^{\it{COP}}$ in each state of $T^a_R$;
\item
The set of initial states $Q_{0R}$ contains elements of the form
$({q_{0R}},{q_0}_{\cop_1},\ldots, {q_0}_{\cop_k},$\\$ {q_{0}}_{\ogsp},
\lambda)$, where $\lambda$ is any element in $2^{\it{COP}}$;
\item
The transition relation $\triangle_R^a$ is defined by
$({q}_{R}',{q'}_{\cop_1},\ldots, {q'}_{\cop_k},\alpha_1,\lambda_1) \in\\
\triangle_{R}^a(({q}_{R},{q}_{\cop_1},\ldots, {q}_{\cop_k},\alpha_1,\lambda_1),
((a_1,\alpha_1,\lambda_1),(a_2,\alpha_2,\lambda_2)))$ if and only if
\begin{itemize}
\item
${q}_{R}' \in \delta_R(q_R,(a_1,a_2))$ and $q'_{\cop_i} =
\delta_{\cop_i}(q_{\cop_i},a_1)$, for $1\leq i \leq k$,
\item
$\alpha_2 \in \delta_\ogsp(\alpha_1,\lambda_1),$ which checks that we
have a run of $A_\ogsp$;
\end{itemize}
\item
The set of accepting states $F_R^a$ contains states of the form
$({q}_{R},{q}_{\cop_1},\ldots, {q}_{\cop_k},$\\$\alpha_1,\lambda_1)$ with
for $1\leq i \leq k$, $q_{cop_i}' \in F_{cop_i}\ \ \mbox{iff}\ \ cop_i
\in \lambda_1$.
\end{itemize}

\noindent
Observe that, since $T_R$ is deterministic weak and the $\omega$-state
properties are represented by deterministic weak automata, the
transducer $T_R^a$ is also deterministic weak.
\\
\newline
\noindent
The initial states of $M^a_\ogsp$ are those of the following form:
$$(w(0),{q_{0}}_\ogsp,\lambda)(w(1),{q_{0}}_\ogsp,\lambda)\cdots$$ where $w(0)w(1)\cdots \in L(A_{S_0})$, and $\lambda$ is any
element of $2^{\it{COP}}$. \\
\newline
The set of accepting states of $M^a_\ogsp$ are those of the following
form:
$$(\Sigma\times q_\ogsp {\times}\it{COP}_i)(\Sigma\times q_\ogsp
\times \it{COP}_i)\cdots$$ where $\it{COP}_i\in 2^{\it{COP}}$ and
$q_\ogsp \in F_\ogsp$. We directly see that the sets of initial and
accepting states can be represented by deterministic weak automata.

\begin{thm}
The B\"uchi regular system $M^a_\ogsp$ has an accepting execution of
the form $\pi^a=\pi^a(0)\pi^a(1)\dots$ if and only if the execution
$\pi=w_0w_1w_2\dots$ of $M$, where $w_i=\Pi_\Sigma(\pi^a(i))$
($\forall i$), does not satisfy $\gsp$.
\end{thm}

\begin{pf}
Follows from the construction above.
\end{pf}

As already mentioned, testing the emptiness of $M^a_\ogsp$ can be done
with the techniques developed in Section
\ref{model-global-regular}. Recall that the definition of the {\em
  greatest simulation relation} over $M^a_\ogsp$ is given by the limit
of the (possibly infinite) decreasing sequence of relations
$Sim_0,Sim_1,\dots$ defined as follows:

\begin{eqnarray}
Sim_0 & = & \{(w_1^a,w_2^a)\} \mid w_1^a{\bar{\times}}w_2^a\in (\Sigma^a\times
\Sigma^a)^\omega \ \wedge \fcop(\Pi_\Sigma(w_1^a))=\fcop(\Pi_\Sigma(w_2^a))\\
Sim_{k+1} & = & Sim_k \cap \{ (w_1^a,w_2^a) \in Sim_k \; \mid \; \\
&&\ \ \ \forall w_3^a .((w_1^a,w_3^a) \in T^a_R 
\Rightarrow \exists w_4^a .(w_2^a,w_4^a) \in T^a_R 
\wedge (w_3^a,w_4^a) \in Sim_k )\} \nonumber, \forall k\in \nats
\end{eqnarray}

A lower approximation of the limit of this sequence can be computed
with the techniques introduced in \cite{BLW04a,Leg07}. In the present
case, the technique requires that each of the $Sim_k$ can be
represented by a deterministic weak automaton. It is easy to see that
$Sim_0$ can be represented by a deterministic weak B\"uchi
automaton. However, the fact that $Sim_k$ is represented by a
deterministic weak B\"uchi automaton does not necessarily imply that
$Sim_{k+1}$ can be represented in the same way. Indeed, building
$Sim_{k+1}$ from $Sim_k$ requires projection operations, and there is
no theoretical guarantee that the resulting automaton can be turned to
a weak deterministic one.

\section{Linear Temporal Properties for Parametric Systems : Parametrization}
\label{temporal-rmc2}

Suppose that we are working with a regular system representing a
parametric system. Global system properties allow to express {\em
  communal temporal properties} of parametric systems, {\emph{i.e.}
  properties such as ``if a process is in a state $s_1$, then finally
  some (possibly different) process will reach a state $s_2$. However,
  global system properties cannot express {\em individual temporal
    properties}, {\emph{i.e.} properties such as ``if the process $i$
    is in a state $s_1$, then finally the process $i$ (the same
    process) will reach a state $s_2$''. Indeed, global system
    properties can only reason on the whole execution of a system,
    while individual temporal properties require to reason on the
    execution of one of the processes.  In this section, we define a
    new class of temporal properties that allows to express individual
    temporal properties of parametric systems.

\subsection{Definitions}
\label{temporal-rmc21}

In our model, an execution of a parametric system is represented by an
infinite sequence of identical length finite words. Each position
in these words corresponds to the state of a process, also called a
{\em local state}, and the infinite sequences of identically
positioned letters in an execution represents a process execution. We
thus use the following notations and definitions.

\begin{defn}
Consider an execution $\pi = w_0w_1w_2w_3\ldots$ of a regular system
$M=(\Sigma, A_{S_0}, T_R)$. The $j$th local projection $\Pi_j(\pi)$ is
the infinite word $w_{0}(j)w_{1}(j)w_{2}(j)\cdots$\,.
\end{defn}

\noindent
Given an execution $\pi = w_0w_1w_2w_3\ldots$ of a parametric system,
the $j$th local projection $\Pi_j(\pi)$ corresponds to the execution
of the $j$th process.

\begin{defn}
Given an alphabet $\Sigma$, a {\em local execution property} is a set
$\lep \subseteq \Sigma^\omega$ that can be represented by a B\"uchi
automaton.
\end{defn}

\noindent
A local execution property $\lep$ is {\em satisfied by an execution
$\pi$ of a parametric system at position $j$}, denoted $\Pi_j(\pi)
\models \lep$, if and only if $\Pi_j(\pi) \in \lep$.\\
\newline
We are now ready to define a logic suited for parametric systems.

\begin{defn}
Given a set of local execution properties
$LEP=\{\lep_1,\ldots\lep_k\}$, a {\em local-oriented system property}
is a set $\losp \subseteq (2^{LEP})^*$, i.e. a set of finite sequences
of subsets of $LEP$, that can be represented by a finite-word
automaton. 
\end{defn}

\noindent
Assume a local-oriented system property $\losp$ defined over $LEP$. An
execution $\pi$ of a parametric system $M$ satisfies $\losp$, denoted
$\pi\models \losp$, if and only if
$\flep(\Pi_1(\pi))\flep(\Pi_2(\pi))\cdots \flep(\Pi_n(\pi)) \in
\losp$, where $n$ is the common length of the words in $\pi$, and
$\flep(\Pi_i(\pi))=\{\lep_i\in LEP \mid \Pi_i(\pi) \models
\lep_i\}$. We say that $M$ satisfies $\losp$, denoted $M\models
\losp$, if and only if all its executions satisfy the property.\\

\begin{figure}
\begin{center}
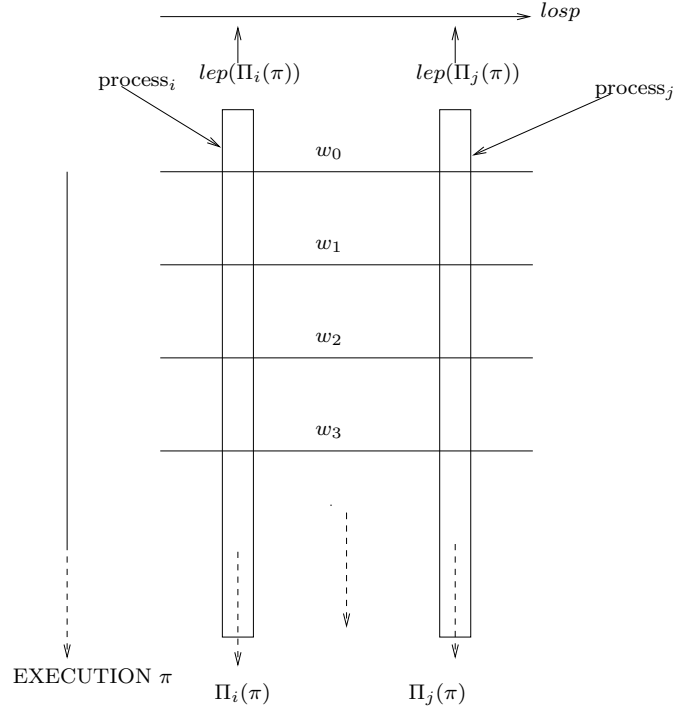
\caption{Local-oriented system properties: an illustration.}
\label{local-figure}
\end{center}
\end{figure}

\noindent
The definition of local-oriented system properties is illustrated in
Figure \ref{local-figure}.

\begin{exmp}
\label{para-one}
Consider the parametric system defined in Example
\ref{example-token}. Given a natural $i$ and a state $N$, the Boolean
proposition $N[i]$ is true if and only if the $i$-th process involved
in the computation (i.e. the one whose state is encoded in the $i$-th
letter of the word describing the global state) is in state $N$. The
fact that whenever a process $i$ is in state $N[i]$, it will
eventually move to state $T[i]$
(${\Box}(N[i]{\Rightarrow}{\diamondsuit}T[i])$ using the well-known
notations for LTL) is a local execution property. That this property
holds for each process ($\forall i
({\Box}(N[i]{\Rightarrow}{\diamondsuit}T[i])$) is then a
local-oriented system property. It is easy to see that this property
is trivially satisfied by the system. Indeed, the transition relation
does not allow for a process to keep the token indefinitely.
\end{exmp}


\subsection{Verification}

Consider a regular system $M=(\Sigma, A_{S_0}, T_R)$ that represents a
parametric system, a set of local execution properties
$LEP=\{\lep_1,\ldots\lep_k\}$, and a local-oriented system property
$\losp$ defined over $LEP$. Suppose that for $1\leq i\leq k$, $lep_i$
is represented by a B\"uchi automaton $A_{\lep_i}=(Q_{\lep_i},\Sigma,
{q_0}_{\lep_i}, \triangle_{\lep_i}, F_{\lep_i})$, which is assumed to
be {\em complete}.
We extend the automata theoretic approach of \cite{VW86} towards a
semi-algorithm to test whether $M$ satisfies $\losp$. Our approach
consists in three successive steps that are the following:

\begin{enumerate}
\item
Computing a deterministic finite-word automaton
$A_\olosp=(Q_\olosp,2^{LEP},$\\$ {q_0}_\olosp, \delta_\olosp,
F_\olosp)$, which is the finite-word automaton accepting the finite
sequences that do not satisfy $\losp$, {\emph{i.e.}} sequences in
$\olosp=(2^{LEP})^*\setminus \losp$;
\item
Building a B\"uchi regular system $M^a_\olosp =
(\Sigma^a, A^a_{S_0}, T^a_{R},F^a)$ whose accepting executions
correspond to those of $M$ that are accepted by $A_\olosp$;
\item
Testing whether $M^a_\olosp$ is empty or not. 
\end{enumerate}

The property $\losp$ being (by definition) representable by a
finite-word automaton, on can always compute the automaton
$A_{\olosp}$. Computing $M^a_\olosp$ is a much harder endeavor for
which we propose the following solution.

For each automaton $A_{\lep_i}$, we assume the existence of a {\em
complete} automaton
$A_{\neg\lep_i}=(Q_{\neg\lep_i}\Sigma,{q_0}_{\neg\lep_i},
\triangle_{\neg\lep_i}, F_{\neg\lep_i})$ whose accepted language is
the complement of the one of $A_{\lep_i}$. Consider an execution
$\pi^a$ of $M^a_\olosp$. Since, {\em a priori}, we do not know which
local execution property will be satisfied by which process, each of
the automata $A_{\lep_i}$ and $A_{\neg\lep_i}$ has to be run in
parallel{\footnote{This can be achieved since the automata are
complete.}} with the local executions of the processes involved in
$\pi$. So, we need to extend the alphabet of $M$ in such a way that
each local state is now also labeling by a state of each of the
$A_{\lep_i}$ and $A_{\neg\lep_i}$. For each $1\leq i\leq k$, running
$A_{\neg\lep_i}$ is necessary since the automaton $A_{\lep_i}$ being
nondeterministic, the fact that it has a nonaccepting run does not
indicate that the corresponding property does not hold.

Furthermore, in each position, each property $\lep_i \in LEP$ might
be satisfied ($A_{\lep_i}$ has an accepting run), or might not be
satisfied ($A_{\neg\lep_i}$ has an accepting run). We make a note of
these facts by also labeling each position by an element of
$2^{LEP}$ corresponding exactly to the properties $\lep_i$ that are
satisfied. This labeling will remain unchanged from position to position and
will enable us to run the automaton $A_\olosp$. The next step is to
check whether there is an execution of $M^a_\olosp$ that is accepting
for suitable automata $A_{\lep_i}$ and $A_{\neg\lep_i}$. Precisely, at
a given position $j$ in the state, the run of the automaton
$A_{\lep_i}$ has to be accepting if $\lep_i \in \flep_j$ and the run
of $A_{\neg\lep_i}$ has to be accepting if $\lep_i \not\in \flep_j$,
where $\flep_j$ is the element of $2^{LEP}$ labeling that position. We
face thus with the problem of checking not one, but several B\"uchi
conditions, {\emph{i.e.} a generalized B\"uchi condition. To do this,
  we use the fact that a generalized B\"uchi automaton has an
  accepting run exactly when it has an accepting run that goes
  sequentially through each of the accepting sets. We now define
  $M^a_\olosp$. The augmented alphabet is $$\Sigma^{a} = \Sigma \times
  \prod_{1\leq i\leq k} Q_{\lep_i} \times \prod_{1\leq i\leq k}
  Q_{\neg\lep_i} \times 2^{LEP}\times 2^{LEP}\times\{reset,
  noreset\}.$$ We thus have two subsets of $LEP$, the second being
  used to remember if suitable automata checking for properties
  $\lep_i$ (or $\neg\lep_i$) have seen an accepting state; the last
  component of the labeling indicates whether the second of these
  subsets has just been reset of not. We denote by $\Pi_\Sigma(w^a)$,
  the word $w\in \Sigma^*$ obtained from $w^a$ by removing all the
  symbols that do not belong to $\Sigma$.

\noindent
An execution $\pi^a=w^a_0w^a_1w^a_2\dots$ of $M^a_\ogsp$ is an infinite
sequence of finite words over $\Sigma^a$ that has to satisfy three
requirements:

\begin{enumerate}
\item
For each $i\,{\geq}\,1$ $(\Pi_\Sigma(w^a_{i-1}),\Pi_\Sigma(w^a_i))\in
T_R$, which ensures that the transitions of $M^a_\ogsp$ are compatible
with the transition relation of $M$;
\item
For each position in a state, the labeling by states of the
$A_{\lep_i}$ form a run of these automata;
\item
The labeling of each position by elements of $2^{LEP}$ stays the same
when moving from one state to the next one.
\end{enumerate}

\noindent
\noindent
We have to build $A^a_{S_0}$, $F^a$, and $T_{R}^a$ in such a way that
the three requirements above are satisfied.

Let $T_R=(Q_R,\Sigma^2, {q_{0R}}, \delta_R, F_R)$. The possibly
nondeterministic transducer
$T^a_{R}=(Q_R^a,(\Sigma^a)^2,q_{0R}^a,\triangle_R^a,F_R^a)$ is built
as follows:

\begin{itemize}
\item Its set of states and accepting states are $Q_{R}^a = Q_R$ and
  $F_{R}^a=F_R$, respectively; its initial state is $q_{0R}^a=
  {q_0}_R$;
\item
The transition relation is defined by (assuming nondeterministic
automata) 
{\small $$\begin{array}{lcl}
(q_{R}^{a})' &\in
\triangle({q}_{R}^a,
(&(a_1,{q_1}_{{\lep_1}}, \ldots,{q_1}_{{\lep_k}},
{q_1}_{{\neg\lep_1}},\ldots,{q_{1}}_{{\neg\lep_k}},
\flep_1,{\flep_F}_1, \rho_1),\\ 
&&(a_2,{q_{2}}_{{\lep_1}}, \ldots,{q_{2}}_{{\lep_k}},
{q_{2}}_{{\neg\lep_1}},\ldots,{q_{2}}_{{\neg\lep_k}},\flep_2,{\flep_F}_2,
\rho_2)))
\end{array}$$}
 if and only if
\begin{itemize}
\item
for $1\leq i \leq k$, $(q_{R}^{a})' \in
\triangle_R^a(q_{R}^a,(a_1,a_2))$ and ${q_{2}}_{{\lep_i}}\in
\delta_{\lep_i}({q_{1}}_{{\lep_i}},a_1)$, ${q_{2}}_{{\neg\lep_i}}\in
\delta_{\neg\lep_i}({q_{1}}_{{\neg\lep_i}},a_1)$,
\item
$\flep_1 = \flep_2$,
\item
if ${\flep_F}_1 = LEP$, then ${\flep_F}_2 = \emptyset$ and $\rho_2
= reset$, or ${\flep_F}_2 = {\flep_F}_1$ and $\rho_2 = noreset$,
otherwise, ${\flep_F}_2 = {\flep_F}_1 \cup \{\lep_i \in \flep_1 \mid q_{{\lep_i}1}
\in F_{\lep_i}\}\cup \{\lep_i \not\in \flep_1\mid q_{{\neg\lep_i}1}
\in F_{\neg\lep_i}\}$ and $\rho_2 = noreset$.
\end{itemize}
Note that at a given position, when all required accepting conditions
have been satisfied, the choice to reset or not is nondeterministic,
which makes it possible to wait until the required acceptance
conditions have been satisfied at each position and then to reset
everywhere simultaneously;
\item
The set of accepting states $F_R^a$ is $ F_R$.\\
\end{itemize}

\noindent
The initial states of $M^a_\ogsp$ are those of the following form:
\[\begin{array}{l}
(w(0),{q_0}_{{\lep_1}},\ldots,{q_0}_{{\lep_k}},
{q_0}_{{\neg\lep_1}},\ldots,{q_{0}}_{\neg\lep_k},
\flep_1,\emptyset,noreset)\\
(w(1),{q_0}_{{\lep_1}},\ldots,{q_{0}}_{{\lep_k}},
{q_0}_{{\neg\lep_1}},\ldots,{q_0}_{{\neg\lep_k}},
\flep_2,\emptyset,noreset)\\ 
\ \ \ \ \ \ \ \ \ \ \ \ \ \ \ \ \ \ \ \ \ \ \ \ \ \ \ \ \ \ \cdots\\
(w(n-1),{q_0}_{{\lep_1}},\ldots,{q_0}_{{\lep_k}},
{q_0}_{{\neg\lep_1}},\ldots,{q_0}_{{\neg\lep_k}},
\flep_n,\emptyset,noreset),
\end{array}\]
where
$w(0)\cdots w(n-1) \in L(A_{S_0})$ and $\flep_1\flep_2\cdots\flep_n \in
\olosp$.\\
\newline
\noindent
The accepting states in the language of the automaton $F^a$ are those
in which for every position the last part $\rho$ of the label is
$reset$, which implies that all relevant automata have seen an
accepting state since the last ``reset''.

\begin{thm}
The B\"uchi regular system $M^a_\olosp$ has an accepting execution of
the form $\pi^a=\pi^a(0)\pi^a(1)\dots$ if and only if the execution
$\pi=w_0w_1w_2\dots$ of $M$, where $w_i=\Pi_\Sigma(\pi^a(i))$
($\forall i$), does not satisfy $\olosp$.
\end{thm}

\begin{pf}
Follows from the construction above.
\end{pf}

The system $M$ being locally-finite, $M^a_\olosp$ is also
locally-finite. We thus have the following result that shows that
checking the emptiness of $M^a_\olosp$ can be reduced to solving the
regular reachability problems.

\begin{prop}
The B\"uchi regular system $M^a_\olosp=(\Sigma^a,A^a_{S_0},T_R^a,F^a)$
is empty if and only if
$$L((T_R^a)^*(A^a_{S_0}) \cap F^a \cap \Pi_{\not=2}((T_R^a)^+ \cap
T_{id}))=\emptyset.$$
\end{prop}

\begin{pf}
Same as Proposition \ref{prop-proof}.
\end{pf}

\section{Boolean Combinations  and Multiple Alternations for Parametric Systems}
\label{bool-liveness}


\subsection{Boolean Combinations}

It is easy to see that one can verify Boolean combinations of global
and local-oriented system properties (each property being a
literal). Indeed, any Boolean combination can be turned into another
combination that only uses the connectors for the disjunction ($\vee$)
and the negation ($\neg$). Properties being defined by finite-word and
B\"uchi automata, one can always compute their negation. Verifying the
disjunction of several properties is direct by definition. 

\subsection{Multiple Alternations for Parametric Systems}

In some situations, it is also interesting to consider properties with
{\em multiple alternations} between local-oriented and global system
properties. By multiple alternations, we mean local-oriented
properties that reference global system properties and vice-versa.  We
will not formally characterize the way alternations can occur, but
rather illustrate the concept with several
examples. Multiple-alternation properties will be specified by
combining the notations introduced in Sections \ref{temporal-rmc11}
and \ref{temporal-rmc21}. The semantics of multiple-alternation
properties easily follows from those notations.\\
\newline
We now propose several examples that illustrate how
multiple-alternation properties can be reduced to properties with a
simple alternation on an augmented system, a problem for which this
paper provided verification procedures. We consider a parametric
system, and assume that each of its processes can be in one of the two
following states ${\lbrace}C,T{\rbrace}$. The following property is a
local-oriented system property:

\begin{equation}
\forall i \Box (C[i] \Rightarrow \Diamond T[i]).
\end{equation}

\noindent
Indeed, we could think that this property is a local oriented system
property. However, due to the presence of the $\exists$ quantifier,
$\Box (C[i] \Rightarrow \Diamond (\exists j\not= i) T[j])$ can
reference several processes and is thus not a local execution
property.

The solution we propose is to reduce the property above to a local
execution property over an augmented system. This is done by
introducing new Boolean variables in the specification of each
process. Those variables can be arbitrarily true or false in any
moment of an execution. Let us go back to our example and assume that
we add to each process a Boolean variable ``a'' that behaves as
described above. We use $a[i]$ to denote that the variable $a$ is true
for the process $i$ in the current state, and $\neg a[i]$ to denote
that it is false{\footnote{When we add a Boolean variable, we extend
    the alphabet on which processes's states are encoded. As an
    example, if the set of states was given by
    $\Sigma={\lbrace}C,T{\rbrace}$ before the variable $a$ is added,
    it becomes $\Sigma_{{\lbrace}a{\rbrace}}= \Sigma\times
    {\lbrace}\neg a, a{\rbrace}= {\lbrace}(C,\neg a),(C,a),(T,\neg
    a),(T,a){\rbrace}$ after the addition occurs. As a consequence,
    any automaton defined over $\Sigma$ must take this extension into
    account, which is done by duplicating each of its transitions. As
    an example, a transition labeled by $T$ is duplicated into two
    transitions, one labeled by $(T,a)$ and the other one by $(T,\neg
    a)$. To not lengthen the presentation, we will assume this
    translation to be implicit, and we write $a[i]$ for $(T,a)[i]\vee
    (C,a)[i]$ and $T$ for $(T,a)[i]\vee (T,\neg a)[i]$.}}. In this
case Property \ref{liv-prop1} can be rewritten as

\begin{eqnarray}
\label{liv-prop1e}
\forall i \Box (C[i] \Rightarrow \Diamond a[i]) \wedge \\
\Box \forall i (a[i] \Leftrightarrow (\exists j\not= i)T[j]) \wedge \\
\Box \forall i (a[i] \vee \neg a[i]).
\end{eqnarray}

\noindent
Clearly, $\phi_1\equiv \forall i \Box (C[i] \Rightarrow \Diamond
a[i])$ is a local-oriented system property, and $\phi_2\equiv \Box
\forall i (a[i] \Leftrightarrow (\exists j\not= i) T[j])$ and
$\phi_3\equiv \Box \forall i (a[i] \vee \neg a[i])$ are global system
properties.\\
\newline
We now give two other illustrating examples. 

\begin{exmp}
Consider the following property:

\begin{equation}
\label{liv-prop2}
\Box (\forall i \Diamond T[i] \wedge \exists j C[j]).
\end{equation}

\noindent
This property cannot be expressed neither by a local-oriented system
property nor by a global system property. The solution is again to
reduce the extended state property to a state property over an
augmented system. We introduce a Boolean variable ``a'' that can be
either true or false in each state. Using variable ``a'', Property
\ref{liv-prop2} can be rewritten as a conjunction of local-oriented
and global system properties.

\begin{eqnarray}
\Box (\forall i a[i] \wedge \exists j C[j]) \wedge\\
\forall i \Box (a[i] {\Rightarrow} \Diamond T[i]) \wedge\\
\Box \forall i (a[i] \vee \neg a[i]).
\end{eqnarray}
\end{exmp}

Of course, we can have several alternations in the same formula. In
such situations, construction has to be applied for each
alternation. Consider the following example.

\begin{exmp}
Consider the following property $\varphi_1$:
\begin{equation}
\label{liv-prop1}
\forall i \Box (C[i] \Rightarrow \Diamond (\exists j\not= i) Buchi_{\varphi}[j]),
\end{equation}
where $Buchi_{\varphi}[j]$ is a {\em B\"uchi modality} which is true
if and only if the $j$-th process satisfies the local execution
property $\varphi$
described by a B\"uchi automaton $Buchi_{\varphi}$.\\
\newline 
Property $\varphi_1$ cannot be expressed neither by a
local-oriented system property nor by a global system property. The
solution is to introduce two Boolean variables ``a'' and $b$. Using
those variables, $\varphi_1$ can be rewritten as the property
$\varphi_2$ defined as follows:

\begin{eqnarray}
\label{liv-prop1e}
\forall i \Box (C[i] \Rightarrow \Diamond a[i]) \wedge \\
\Box \forall i (a[i] \Leftrightarrow (\exists j\not= i)b[j]) \wedge \\
\forall i \Box (b[i] \Rightarrow Buchi_{\varphi}[i]) \wedge \\
\Box \forall i (a[i] \vee \neg a[i]) \wedge \\
\Box \forall i (b[i] \vee \neg b[i]).
\end{eqnarray}

\noindent
By observing that $\forall i \Box (b[i] \Rightarrow
Buchi_{\varphi}[i])$ is a local-oriented property (The set of
executions that satisfy $b$ can easily be described with a B\"uchi
automaton), we conclude that $\varphi_2$ is a Boolean combination of
local-oriented and global system properties.
\end{exmp}

There are also alternations that we have not been able to handle. As
an example, we cannot treat a property that has two free-variables or
a second order variable under the scope of a temporal LTL
operator. Such an observation was made for a similar logic in
\cite{AJNdS04,AJNS04}.

\section{Related Work on Verifying Temporal Properties in ($\omega$-)Regular Model Checking}
\label{related}

The problem of verifying linear temporal properties in the framework
of regular model checking has been first addressed in
\cite{BJNT00,PS00,Sha01}. However, the treatment of this problem in
these papers was preliminary and somewhat adhoc for very particular
kinds of properties of parametric systems.

In \cite{AJNdS04,AJNS04}, Abdulla et al. independently{\footnote{The
    approach in \cite{AJNdS04,AJNS04} has been proposed in the same
    period of time as our early work\,\cite{BLW04b}, whose present
    paper is an extension of.}}  proposed an approach based on a
specification logic called LTL(MSO), which combines the monadic second
order logic MSO and the linear temporal logic LTL. Properties written
in the LTL(MSO) logic are local-oriented system properties, where the
local system properties are LTL properties that can make assumptions
on the executions of the other processes up to some restrictions. The
LTL(MSO) logic has been designed for parametric systems and is
not suited (and sometimes not powerful enough) to express very simple
properties of many other interesting classes of systems such as
systems with integer variables (when considering a non-unary
encoding). The verification procedure in \cite{AJNdS04,AJNS04} is only
dedicated to regular systems that are locally-finite and the
$\omega-$regular framework is not considered. Finally, unlike our
local-oriented properties, the LTL(MSO) logic cannot be used to
express properties which are Boolean combinations of properties
written in logics that are more expressive/concise than LTL
(e.g. PTL\,\cite{GO03,LPZ85}, ETL\,\cite{Wol82}, or
$\mu$TL\,\cite{Var88}).

In \cite{VSVA05}, Agha et al. proposed to use learning-based
algorithms\,\cite{Ang87} to verify global system properties of regular
systems. The technique they proposed relies on the computation of
several fixed point operators which are used to test whether a B\"uchi
regular system is empty or not. The use of learning algorithms to make
fixed point computation terminating requires to enrich the systems
with two extra variables. This is a clear restriction since it is
known that there are many systems for which the set of reachable
states is regular before the variables have been introduced, but not
after.
The work in \cite{VSVA05} also lacks of a clear description of the
encoding of linear temporal properties in the regular framework, which
is one of the main contribution of our work. Finally, we mention that
\cite{VSVA05} does not consider the $\omega-$regular framework.

\section{Conclusion and Future Work}
\label{conclu}


We have presented a general framework for specifying and verifying a
large class of linear temporal properties for systems represented in
the ($\omega$)-regular model checking framework. The verification
techniques we provide are based on reductions to the
($\omega$-)reachability problems.  

Our objective was not performances evaluation. A next step will thus
be to implement our constructions in several regular model checking
tools (e.g. T(O)RMC\,\cite{Leg08}, LEVER\,\cite{VV06}, or
RMC\,\cite{RMC}) and compare the performances. Another direction for
future work is to extend our results to the verification of
computational tree logics properties. It would also be of interest to
propose criteria to check whether the extrapolation of the simulation
with the technique of \cite{BLW03,BLW04a,Leg07} is precise. Developing
a methodology to decide whether FIFO-Queue and pushdown systems are
locally-finite is another topic of interests. We would also like to
give a formal characterization of what are the allowed alternations
between local-oriented and global system properties.



\section*{Acknowledgement}

We thank Julien d'Orso, Marcus Nilsson, and Mayank Saksena for
answering many email questions on their work.  \bibliography{paper}
\bibliographystyle{alpha}
\end{document}